\pdfoutput=1
\documentclass[12pt,a4paper]{article}          % twocolumn
\usepackage{authblk}
\usepackage[sort&compress,numbers]{natbib}
\usepackage{graphicx}
\usepackage{url}
\usepackage[version=3]{mhchem} % Formula subscripts using \ce{}
\usepackage{chemfig}
\usepackage{amssymb}
\usepackage{pgfplots}

\pgfplotsset{width=8cm,compat=1.7}
\pgfplotsset{every axis label/.append style={font=\sffamily\small},
             tick label style = {font=\footnotesize},}

\setdoublesep{0.3em}             % Chemdraw: 'Bond Spacing'
\setatomsep{2.5em}               %           'Fixed Length'
\setbondoffset{1.6pt}            %           'Margin Width'
\newcommand{\bondwidth}{0.6pt}   %           'Line Width'
\setbondstyle{line width = \bondwidth}

\begin{document}

% \title{Octet rule based generation of virtual molecules}
\title{The octet rule in chemical space: Generating virtual molecules}
\makeatletter
\def\blfootnote{\xdef\@thefnmark{}\@footnotetext}
\makeatother
% \title{The octet rule in chemical space.}
% \title{Generating virtual molecules.}

% \subtitle{Do you have a subtitle?\\ If so, write it here}

\author{Rafel Israels}
\author{Astrid Maa\ss}
\author{Jan Hamaekers}
\affil{Fraunhofer Institute for Algorithms and Scientific Computing SCAI, Schloss Birlinghoven, D-53754 Sankt Augustin}
\maketitle
\blfootnote{The final publication is available at Springer via \url{https://doi.org/10.1007/s11030-017-9775-2}}

\begin{abstract}

We present a generator of virtual molecules that selects valid chemistry on the basis of the octet rule. Also, we introduce a mesomer group key that allows a fast detection of duplicates in the generated structures.

Compared to existing approaches, our model is simpler and faster, generates new chemistry and avoids invalid chemistry. Its versatility is illustrated by the correct generation of molecules containing third-row elements and a surprisingly adept handling of complex boron chemistry.

Without any empirical parameters, our model is designed to be valid also in unexplored regions of chemical space. One first unexpected finding is the high prevalence of dipolar structures among generated molecules.

%\keywords{chemical space, virtual chemistry}
\end{abstract}

\section{Introduction}
\label{intro}

Chemical space can be defined as the search space of all chemically relevant entities. It includes the universe of solid state materials such as alloys and semiconductors that are the focus of materials science. In organic chemistry, the focus of this contribution, chemical space is rather understood to include all distinct molecules. This space is being explored by a number of research groups, including efforts to estimate its size \cite{Reymond}, finding ways to enumerate its elements \cite{vanDeursen} and/or to determine distances between them \cite{vonLilienfeld}. One motivation for this type of research is the declining efficiency of product development in chemistry \cite{Scannell2012}. At first glance, a likely cause for this decline could be that most relevant molecules have been identified after 100 years of chemical research.

A notable project in this field of research is the production of the GDB libraries by the Reymond group. These libraries consist of all relevant molecules that contain carbon, nitrogen, oxygen and hydrogen with, e.g., GDB-11 defined as all such molecules that contain up to 11 non-hydrogen atoms \cite{Reymond, GDB1}. GDB-11 contains $2.6 \times 10^7$ entries; the latest contribution is GDB-17 with $1.7 \times 10^{11}$ entries \cite{GDB2}. In comparison, the number of known structures seems low. For example, the PUBCHEM database aims to collect all known molecules and contains $ \sim 2 \times 10^8$ entries \cite{PUBCHEM}. The commercial CAS database has a similar size \cite{CAS}. This comparison seems to indicate that only a small fraction of chemical space has been explored, thereby contradicting its exhaustion as a cause of declined efficiency in product development.

Another potential cause could be that product development suffers from a selection bias. Because researchers tend to search in well-known regions of chemical space, large areas may have remained unexplored. This is where virtual molecule generators may be expected to be of value: software tools that generate chemical structures based on explicit rules. Libraries that are produced with a virtual molecule generator may be expected to be a good and unbiased start for any product development.

Virtual generators are routinely used in chemical analysis to produce lists of candidates that match spectra obtained from, e.g., NMR or mass spectrometry \cite{hamdalla2013biosm,peironcely2011understanding}. The commercial generator MOLGEN is one such product \cite{gugisch2000molgen}. The open-source generator OMG was developed by Peironcely et al \cite{Peironcely2012} in 2012. It is based on the graph generating tool NAUTY/GENG developed by McKay \cite{mcKay2014} and generates structures by permuting atom types on graph nodes and bond orders on graph segments. The subsequent selection of valid chemistry from all generated structures is based on a combination of simple valency rules with a dictionary of known exceptions that is retrieved from the Chemistry Development Kit (CDK) \cite{Steinbeck2003}.

By contrast, the use of virtual generators in product development remains less established. Most work in this area is in the field of materials science \cite{curtarolo2013,armiento2011}. Three projects focusing on organic chemistry all address the search for photovoltaic molecules \cite{oboyle2011,hachmann2014,cole2014}. In these contributions the virtual chemistry is generated as a variation of existing chemistry and based on intimate chemical expertise on the addressed problem. An example of using an unbiased virtual generator to explore new regions of chemical space is the publication by Husch and Korth \cite{Husch2015}, describing the search for Li-air battery electrolytes. The authors first describe how they selected DMSO as the best candidate from (at the time) 67 million entries in the PUBCHEM database. This result confirms the validity of their approach, but is nevertheless disappointing as DMSO was the high-performing benchmark that the study hoped to replace. The authors then applied the same approach to sets of virtual molecules that had been generated with MOLGEN. However, no promising candidates could be identified among these candidates.

A number of generators have been described that are based on different concepts. Korth and Grimme reduce selection bias to a minimum with a generator that is based on the random generation of atom coordinates, followed by a quantum chemical calculation to select stable and valid chemistry \cite{korth2009}. The LSD program developed by Nuzillard generates structures that match spectroscopic information, based on a logical combination of constraints that are obtained, e.g., from 2D NMR data \cite{nuzillard2003}. The ACSESS framework produces representative and diverse subsets of larger, e.g., enumerated libraries by a stochastic search \cite{rupakheti2015,virshup2013}. The SCUBIDOO database has been generated by applying a set of 58 well-defined reactions to a set of 18561 building blocks and has the advantage that synthetic accessibility of entries is explicitly addressed \cite{chevillard2015}. A promising approach by Kayala and Baldi calculates chemical reactions as combinations of electron movements, with the probability of individual movements inferred from known chemistry by an artificial intelligence module \cite{Kayala2011,kayala2012}.

In pharmaceutical product development, a target is to generate structures that are related to existing lead structures and/or may be expected to have a favorable interaction at the target site. \cite{Hoksza2014} For such applications it is critical to have some measure of distance in chemical space, so that in the most ideal case test structures can be selected that are evenly spaced around a group of known lead structures. A recent review by von Lilienfeld \cite{vonLilienfeld} describes the difficulties involved in defining such distances. Promising approaches are described in two recent contributions \cite{vonLilienfeld2015fourier,desandip2016}.

In this contribution we present a virtual generator that is based on the octet rule. It is similar to OMG, but instead of accepting molecules according to predefined valences of atom types, we accept all structures that obey the octet rule. We compare our new generator to OMG and to GDB-11 by generating sets of molecules containing up to 9 non-hydrogen atoms. The comparison is based on two new keys that allow a consistent treatment of mesomeric and tautomeric structures. For the smallest molecules in this set we validate the generated structures by quantum chemical calculations. Finally, we explore the general validity of our model by applying it to boron chemistry and third-row (period) elements.

The remainder of this paper is organized as follows. In section \ref{methods} we describe the molecule-generating algorithm and its handling of duplicates. This includes a discussion on automorphic, mesomeric and tautomeric structures and a description of the keys that we use to distinguish them. The results section starts with a detailed comparison to OMG and GDB-11, followed by an investigation of performance and a validity check. We conclude the result section with a selection of generated molecules that illustrate both the versatility and the limits of our model. In section \ref{conclusion} we summarize our main findings and present an outlook on future work.

\section{Methods}
\label{methods}
\subsection{Generator}
\label{generator}
\begin{figure}[!h]
\center
\includegraphics[width=0.9\textwidth]{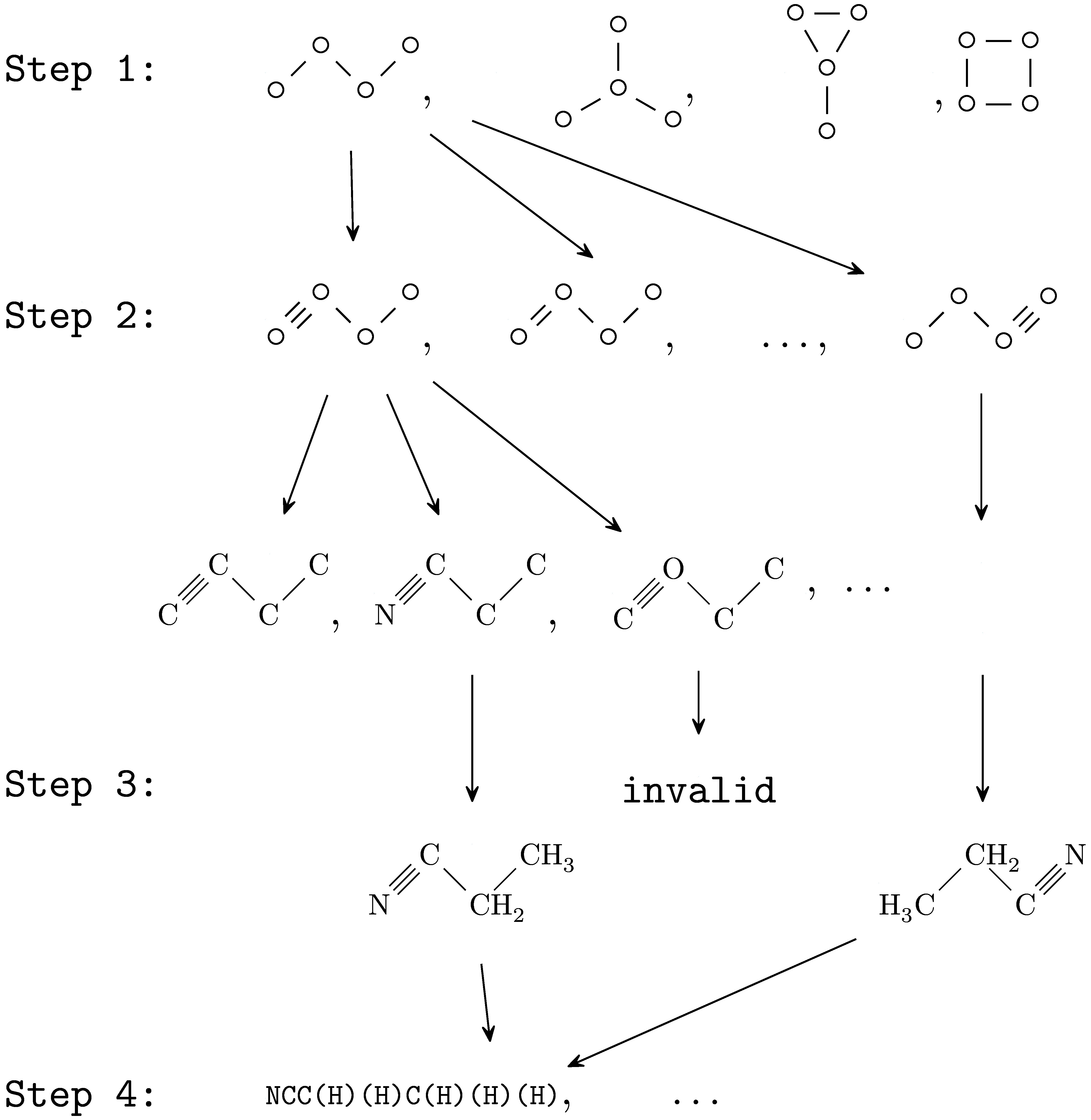}

\caption{Schematic overview of the four steps involved in our generator model: graph generation, bond and atom type permutation, validity selection and duplicate filtering.}

\label{FigureSchema}

\end{figure}

As depicted in Figure \ref{FigureSchema}, our generator is similar to OMG: in a first step we use the GENG/NAUTY tool to generate all unique graphs of a given size, in a second step we permute bond orders on the segments and non-hydrogen element types on the nodes of each graph and in a third step we select valid chemistry. The novelty of our approach is the implementation of the third step and the addition of a fourth step in which we exclude duplicates. This filtering for unique molecules in the final fourth step is described below. First we describe how we select valid chemistry.

In OMG any structure is valid in which all atoms have typical valences like four for a carbon, three for a nitrogen and two for an oxygen atom. This simple rule is completed by a dictionary of non-typical but known-valid atom environments that is being maintained in the Chemistry Development Kit (CDK).

In contrast, we define valid chemistry as molecules in which all
electrons are located in a complete shell. This approach is based on the octet rule,  stating that atoms of the main-group elements tend
to combine in such a way that each atom has eight electrons in its
valence shell, giving it the same electronic configuration as a noble
gas \cite{wikiOctet}. Although a simple model without general validity, the octet rule is the next-best improvement on the assumption of fixed valences. In our approach it replaces the CDK-listed atom-environments.

In our implementation of the octet rule we accept only structures for which the following equation holds:

\begin{align}
\label{eqOctet}
\sum_i \left( v_i + b_i - x_i \right) = \sum_i q_i =0.
\end{align}

In this equation the summation runs over the atoms of a molecule, with $x_i$ the number of electrons that atom $i$ accommodates in its outer shell: $x_i = 2$ for hydrogen and $x_i = 8$ for all other atom types. In our model these outer shells are exactly filled for each atom, so that the formal charge on an atom is calculated as $q = v + b - x$ with $v_i$ the number of valence electrons provided by the $i^\mathrm{th}$ atom and $b_i$ the number of  bonds in which atom i participates. Here we need to count double bonds twice, i.e., we define $b_i$ more precisely as half the number of electrons in bonds comprising the $i^\mathrm{th}$ atom. The summation over all formal atom charges gives the net charge of the molecule, which is set to zero in this publication.

Equation \ref{eqOctet} assists the performance of the algorithm in the second step, by skipping bond permutations with too high a number of bond electrons.
 In the third step of our algorithm, the octet rule validity check is
performed on each backbone, where we define a backbone as one
permutation of bond orders and non-hydrogen atom types for one graph.
 Note that equation \ref{eqOctet} is a necessary but not a sufficient
condition for each electron to be in a complete shell. In the
implementation of our validity check we use this equation to calculate
the total number of hydrogen atoms. Based on a simple heuristic we then
distribute these hydrogen atoms on the backbone atoms and check that we have a
 complete chemical structure that obeys the octet rule. It is not
obvious that such a valid structure must exist for each backbone.

\subsection{Duplicates and keys}
\label{duplicates}
The third step of our generator produces a list of structures in the form of a list of smiles strings \cite{weininger1970smiles}. Each of these strings represents one molecule, but the reverse is not true: one molecule can be represented by more than one smiles string. This means that our list may contain duplicate representations of molecules. In order to generate a list of distinct molecules we need to exclude these duplicates.

There are three kinds of duplicates that need to be considered: those arising from graph automorphism, from mesomerism and from tautomerism. Two structures that are either
automorphic or mesomeric to each other are duplicate representations of one
distinct mol\-e\-cule. In this publication we consider two tautomeric structures to represent two distinct molecules.

\begin{figure*}[!h]
\center
\includegraphics[width=0.9\textwidth]{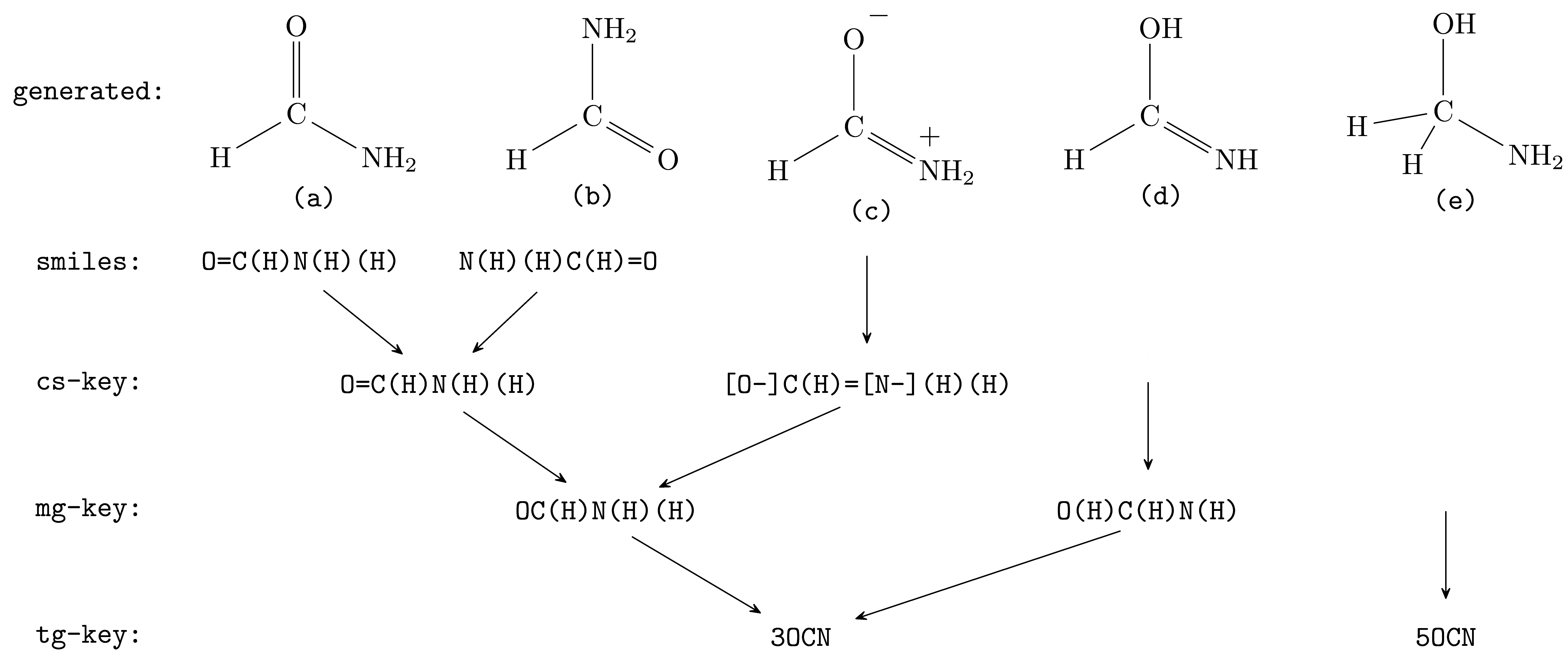}

\caption{Illustration of the different forms of duplicates and the keys that help to distinguish them: the smiles codes for two different automorphic representations of formamide (a,b) share the same cs-key. Also the mesomeric structure (c) still represents formamide and it shares the same mg-key. The tautomeric iminol form of formamide is a different molecule with a different cs-key and mg-key, but it shares the same tg-key with formamide. Methanolamin (e) is a different molecule, neither a tautomer nor an isomer of formamide. }

\label{FigureKeys}

\end{figure*}

We consider two structures automorphic when their nodes and edges can be mapped onto each other without breaking a bond. This is the formal definition for the well known problem that a molecular descriptor depends on the order in which the atoms are considered, so that for example both \ce{HOCH2CH3} and \ce{CH3CH2OH} are equally valid descriptions of ethanol. Two structures are mesomeric when they differ only with respect to the location of electrons. As an example we show automorphic and mesomeric representations of the molecule formamide in Figure \ref{FigureKeys} a, b and c.

Our definition of automorphic and mesomeric structures seems in agreement with definitions \cite{iupacgold} as issued by the International Union of Pure and Applied Chemistry (IUPAC). The case of tautomerism is not so clear though.

IUPAC defines tautomerism as: ``an isomerism of the general form
\[\ce{G-X-Y=Z <-> X=Y-Z-G}\] where the isomers (called tautomers) are
readily interconvertible'' \cite{iupacgold}. Including the ready interconversion in this
definition makes tautomerism to a phenomenon that depends on timescales and conditions and is finally determined by experiment. When comparing and analyzing virtual molecules, as we do here, such a dependence on experimental verification is problematic. Therefore, we adopt a generalized definition and consider any two structures
tautomeric that differ only with respect to the location of protons and
electrons. According to this definition, ethenol is tautomeric to ethanal and 1-hex\-ene is tautomeric to 2-hexene, regardless of the interconversion time-scale. Also, we thus define the zwitterionic forms of amino acids to be tautomers of the nonionic forms and tautomerism to be a transitive relation: if both pairs 1-hexene/2-hexene and 2-hexene/3-hexene are
tautomers, then also 1-hexene and 3-hexene are tautomers.

The most widely used framework to address tautomerism and mesomerism is
the IUPAC International Chemical Identifier (InChI), which also includes
 the definition of a generalized InChIKey \cite{INCHI}. This key is
designed to
allow for easy web searches of chemical compounds and may be expected to
 uniquely identify molecules. It considers tautomers to be different
representations of one molecule, but only if the interconversion is
fast: formamide and
its tautomeric iminol (shown in Figure \ref{FigureKeys} (d))
share the same InChIKey, whereas the tautomers ethenol and ethanal have
different keys. This seems unnecessarily complicated when comparing
millions of virtual structures. In this publication we use the above
generalized definition of tautomerism and define two tautomeric
structures to be representations of two different distinct molecules.

In order to analyze the various forms of duplicates we calculate keys that represent groups of structures. For each of the smiles strings produced in step three we calculate a canonical smiles key (cs-key). This calculation is based on a canonicalization of the node indices, a well-described problem in graph theory, for which many algorithms have
been developed. The cs-key uniquely labels a group of structures that are automorphic to each other. Furthermore, if we calculate $m$ distinct cs-keys for a list of $n$ different smiles strings, we define the automorphism duplication factor $n/m$ as the average number of smiles strings produced per unique cs-key. We will use this duplication factor when discussing the efficiency of our algorithm.

Similarly, for each cs-key we calculate a mesomeric group key (mg-key) that uniquely labels a group of structures that are mesomeric to each other and for each mg-key we calculate a tautomeric group key (tg-key) that labels groups of tautomeric structures. As stated above, in this work a list of unique mg-keys is considered to be a list of unique molecules. We do not need the tg-key to exclude duplicates but we will use this key to analyze the generated molecules. Our use of group keys leads to a wording that
may sound confusing at first: the mesomeric structures of a molecule are
 not distinguished by mesomeric keys. Rather, they share one and the
same mesomer group key and are distinguished by their respective
canonical smiles keys.

Our mesomer group key is constructed as a cs-key without bond order information. Our tautomer group key is a cs-key with neither bond orders nor hydrogen atoms, to which the number of hydrogen atoms is added as a prefix. Examples of the newly defined keys can be found in Figure \ref{FigureKeys}.

\section{Results}
\label{results}

\subsection{Comparison to OMG and GDB-11}
\label{comparison}
In this section we test our generator on sets of small molecules that contain C, N, O and H-atoms only. This restriction is not due to a limitation of our model -- we choose it to allow a comparison to the well-known GDB-11 database \cite{GDB1,GDB2}.

We use the term gdb-x to denote a set of all such molecules of size x, with x defined as the number of non-hydrogen atoms. For example, the (lower case) term gdb-6 denotes a set of molecules with six non-hydrogen atoms. The (upper case) GDB-11 is reserved for the database of molecules as generated and published by Reymond, comprising eleven subsets gdb-1 up to gdb-11.

We have generated the sets gdb-1 up to gdb-9 with our  model and compare them with the corresponding sets extracted from GDB-11 and obtained with OMG. The molecular formulas that OMG requires as input were generated with our software and we used Open Babel \cite{obabel} to generate smiles strings from the sdf files that OMG produces.

Before going into more detail, the comparison can be summarized as follows: our model is comparable to OMG with differences in three well-defined classes of molecules. In comparison to both OMG and our model, GDB-11 contains roughly thousand times fewer entries, due to a focus on pharmacological molecules.

\begin{figure}[!h]
\center
\includegraphics[width=0.6\textwidth]{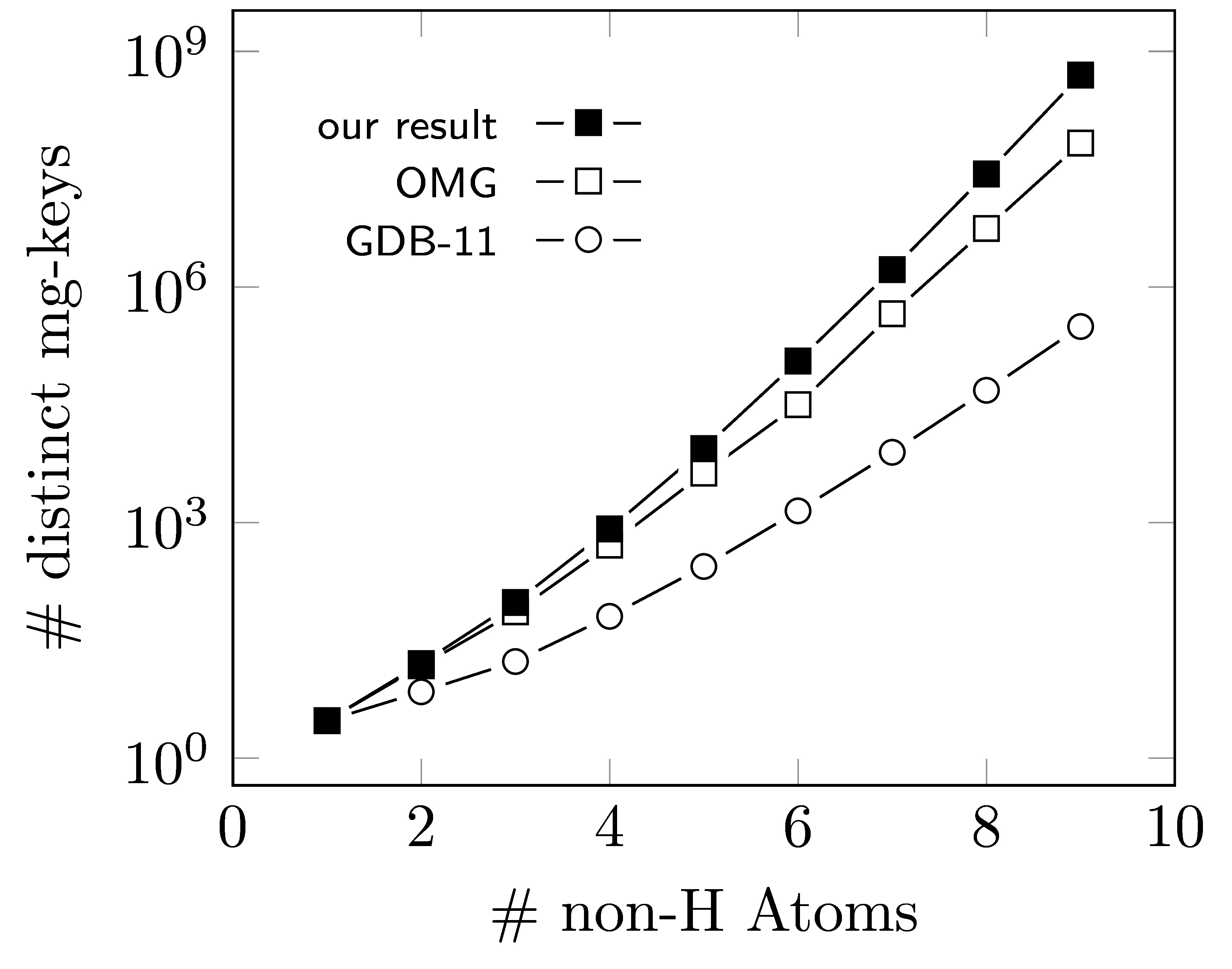}

\caption{Number of unique molecules generated with our software ($\blacksquare$), as compared to those produced with OMG ($\square$) or contained in GDB-11 ($\circ$).}

\label{FigureNumMolecules}
\end{figure}

As discussed in section \ref{methods}, the comparison is based on mesomeric group keys: for each generated structure we calculate the corresponding mg-key and we plot in Figure \ref{FigureNumMolecules} the number of unique mg-keys. To ensure a fair comparison we use the exact same procedure on the sets that we generate with OMG or extract from GDB-11 and thereby check these sets for potential duplicates too. We do find a relatively high number of duplicates in the OMG generated list. For example, gdb-6 as generated by OMG contains 63,295 smiles strings that represent 31,870 distinct mg-keys, on average two mesomeric structures for each distinct molecule. In contrast we do not find a single duplicate in GDB-11.

\begin{figure}[!h]
\center
\includegraphics[width=0.8\textwidth]{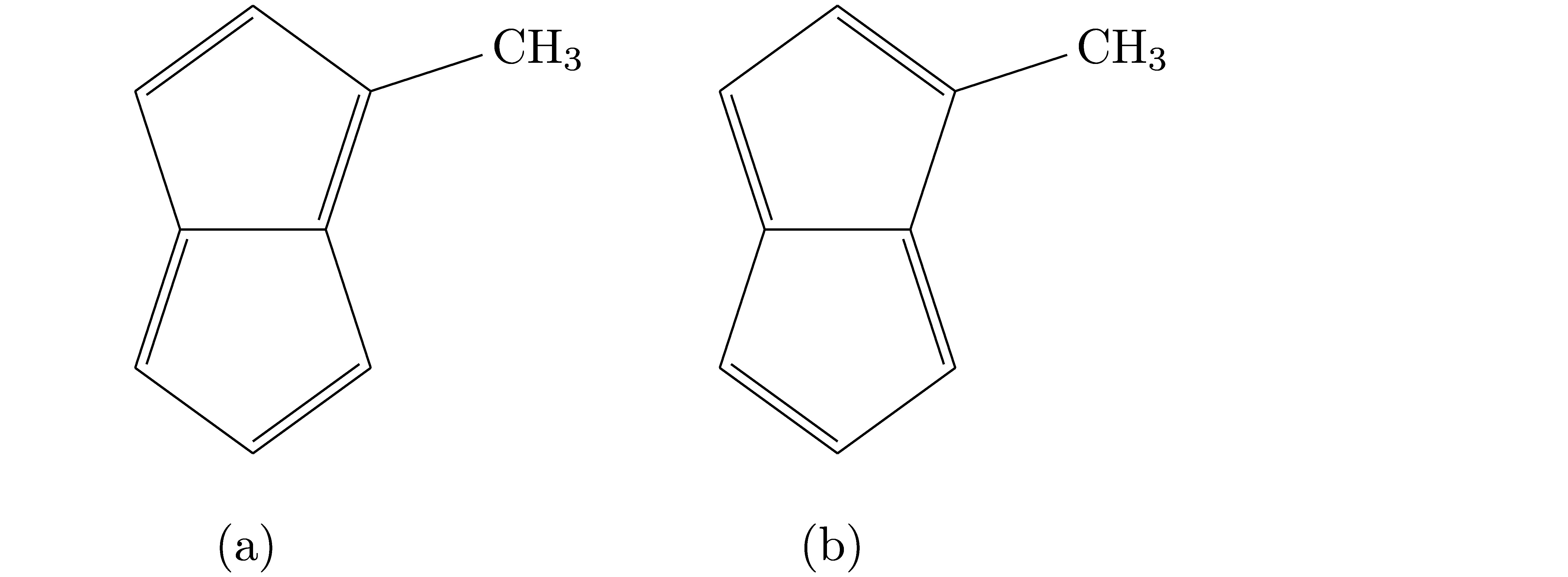}

\caption{Two antiaromatic forms of 1-methylpentalene that are correctly included in the GDB-11 and for which our model includes only one, falsely discarding the other as a ``mesomeric duplicate''. }
\label{FigurePentalene}
\end{figure}

The search for duplicates in GDB-11 does reveal a minor shortcoming of our approach: we calculate one mg-key for pairs of antiaromatic structures, discarding one member of such a pair as ``mesomeric duplicate'' where Reymond correctly includes both. We find 24 such structures, one being methylpentalene, as shown in Figure \ref{FigurePentalene}. All in all, the high degree of agreement of our approach with GDB-11 is an indication for the quality of both approaches.

In Figure \ref{FigureNumMolecules} the solid squares indicate the number of molecules that we generate. Our gdb-1 set consists of the expected molecules methane, ammonia and water. The number of molecules grows strongly with size and reaches half a billion
distinct structures in the gdb-9 set. The open squares and circles indicate the number of molecules that OMG (squares) and Reymond (circles)
generate in agreement with our results. The logarithmic scale that we use to compare results over eight orders of magnitude can be misleading: OMG generates well over a hundred times more molecules than contained in GDB-11 and with our model we generate a further five times more.

The intention of GDB-11 is not to generate all molecules, but rather to focus on pharmacological structures. Accordingly, a number of filters have been applied as described by Fink and Reymond \cite{GDB2}. This explains why GDB-11 contains a subset of the molecules that we generate. For example, the 79 molecules that are missing as compared to our gdb-3
list include carbon dioxide and 48 further molecules
that contain less than two carbon atoms. Also, six missing ring structures and three missing non-cyclic imines can be attributed to explicit
filtering rules. All in all GDB-11 still contains over 400\;000 molecules in the sets gdb-1 to gdb-9 and it is reassuring that -- except for the 24 antiaromatic duplicates -- we find every single one also in our lists.

\begin{figure}[!h]

\center
\includegraphics[width=0.6\textwidth]{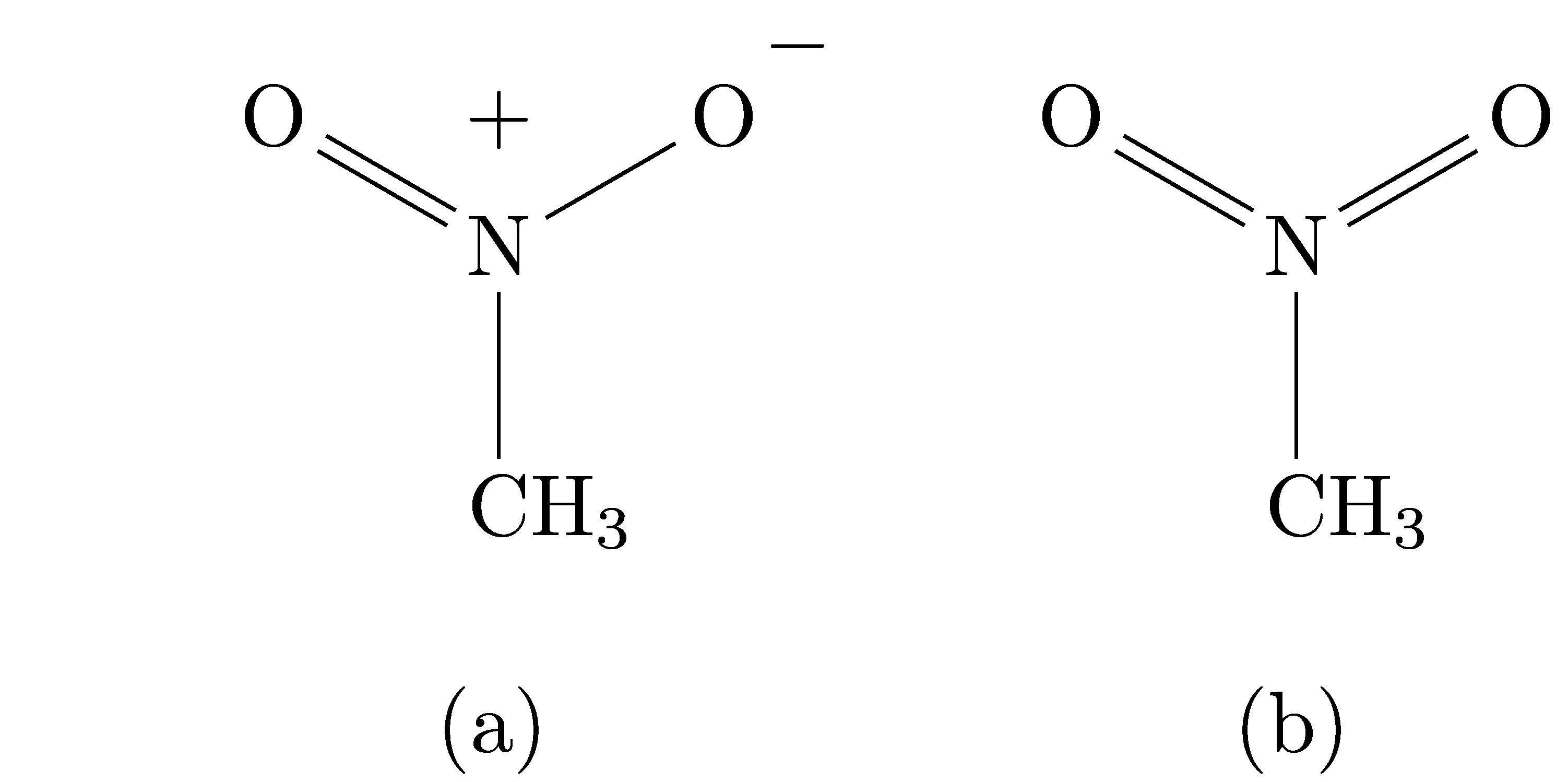}

\caption{Nitromethane depicted as Lewis structure (a) or
according to OMG (b). }
\label{FigureNitromethane}
\end{figure}

Comparing our results to those generated by OMG, we see that the two
tools are in complete agreement on gdb-1 with three molecules ammonia,
water and
methane. First deviations show up in gdb-2 with OMG not generating
carbon monoxide and in gdb-3 with 23 structures missing in the OMG
generated list, including real chemicals like ozone, nitrous acid,
hydraxoic
acid, fulminic acid, diazomethane and methylisocyanide. The common
feature of these structures is
 that they contain dipolar bonds. On the molecules without dipolar
bonds, the two tools are in
 complete agreement up to and including gdb-9.

\begin{figure}[!h]
\center
\includegraphics[width=0.6\textwidth]{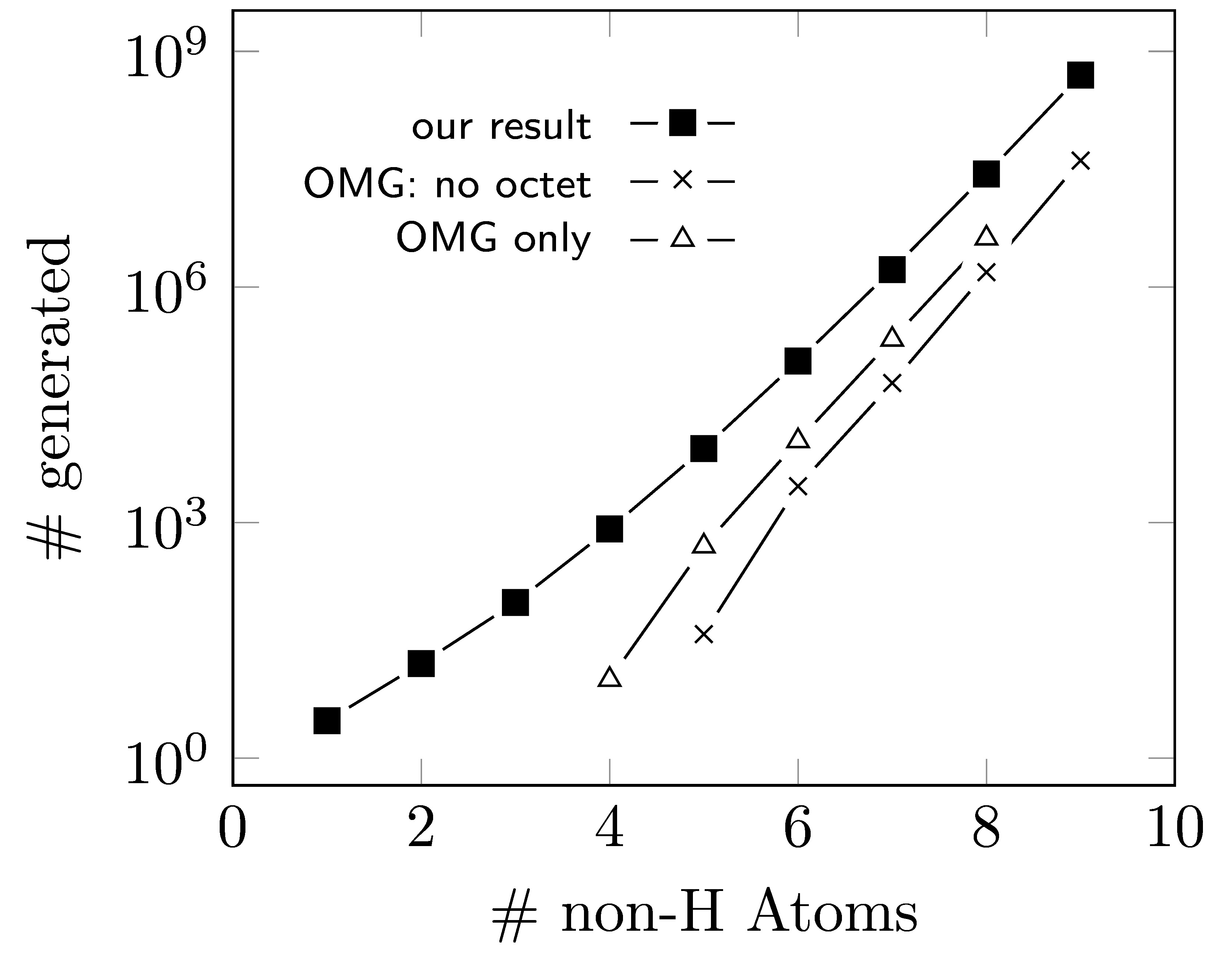}

\caption{Number of octet-rule-violating molecules generated by OMG
($\times$) and number of OMG-generated octet-rule-complying
molecules that we fail to generate ($\triangle$). As a comparison we repeat from Figure \ref{FigureNumMolecules} the number of molecules we generate ($\blacksquare$).}
\label{FigureDiscrepancies}
\end{figure}

In this context the term dipolar bond is used to describe a conceptual charge separation \cite{iupacgold}, not necessarily concomitant with an actual charge separation. An informal more intuitive definition is that the atoms participating in dipolar bonds have atypical valencies: valencies that deviate from
the standard of four bonds for carbon, three for nitrogen and
two for oxygen. Not all such structures are missing in the OMG lists.
 One example is nitromethane, for which we show both our and OMG's
representation in Figure \ref{FigureNitromethane}. In our representation
 one oxygen has
an untypical valency of one, nitrogen has an untypical valency of four and the charge separation is a direct consequence of the octet rule and
explicitly included. The OMG representation is based on a
 non-typical valence of five for nitrogen and does not include a charge
separation.

Among the molecules with untypical valencies and dipolar bonds, we
find three classes of discrepancies between our generator and OMG. The
first class is indicated by triangles in Figure
\ref{FigureDiscrepancies} and comprises missed tautomers. To be precise:
 the triangles indicate the number of OMG generated molecules that we
fail to generate. In all cases we do find the corresponding tautomeric
group key in our result: for each of these missed molecules we generate
at least one of its tautomers. In comparison to the total number of
structures, again indicated as solid squares, it seems that for one in
ten molecules we miss a tautomer that is generated by OMG.

The second class of discrepancies are those molecules that OMG fails
to generate. As described above, carbon monoxide and ozone are two
representatives of this class. The magnitude of this discrepancy is
given by the difference between the two curves in Figure
\ref{FigureNumMolecules} and seems significant for larger molecules: in gdb-8 OMG generates on average only one out of four molecules we
 generate.

\begin{figure}[!h]

\includegraphics[width=0.8\textwidth]{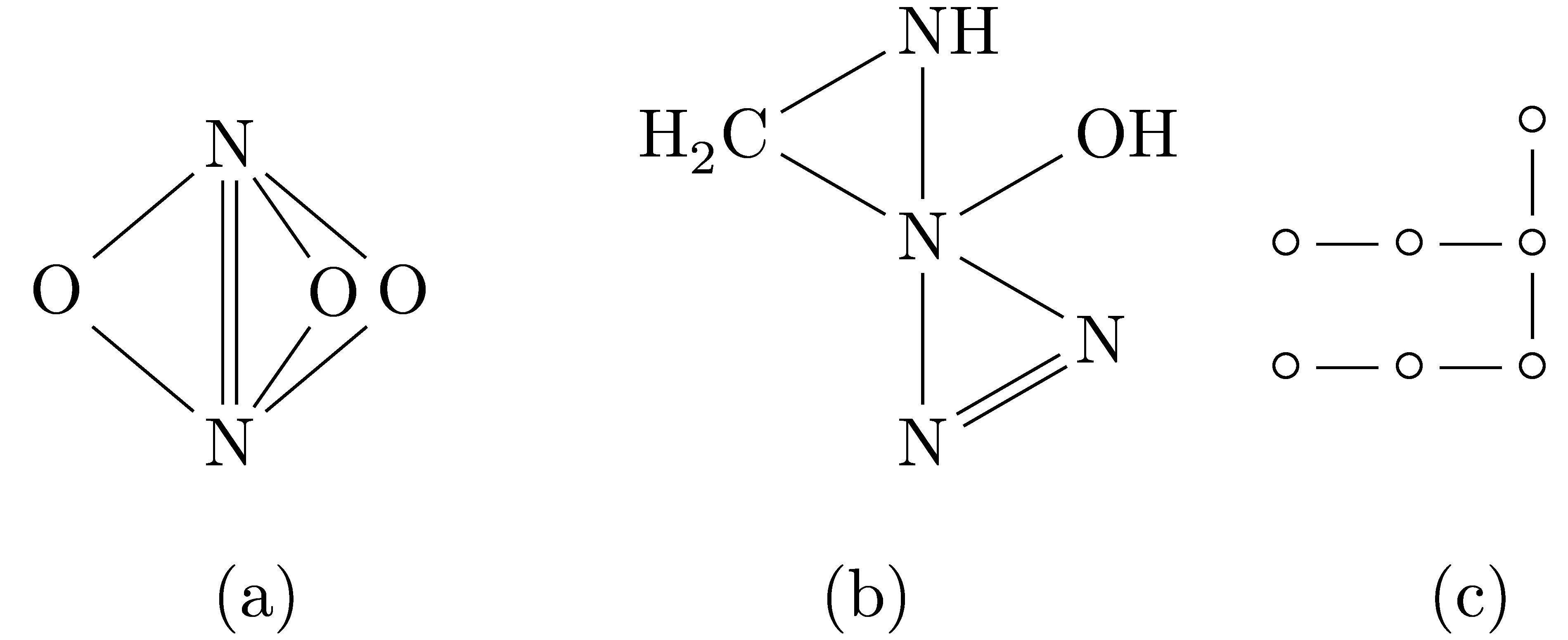}
\caption{Examples of structures that are generated by OMG in violation of the octet rule (a,b) and the graph corresponding to 3-methyl-hexane, which has no symmetry (c): no two nodes are equivalent. }

\label{FigureNoOctet}
\end{figure}

The third class of discrepancies comprises the OMG-generated
 molecules that violate the octet rule. These are indicated by crosses
in Figure \ref{FigureDiscrepancies} and it is obvious that -- by design -- these are not
contained in our result. All structures in this class are either
 structures in which one atom has more than four neighbors, or
structures
 for which equation \ref{eqOctet} indicates a surplus of bonds. Typical
examples are shown in Figure
\ref{FigureNoOctet} (a) and (b). The structure shown in (a) is obviously unlikely because of its complicated multi-ring nature, but its violation of the octet rule is the following: Based on equation \ref{eqOctet} we compare $2\times5 + 3\times 6 = 28$ valence electrons and\hspace{1em}$8\times 2 = 16$ bond electrons with\hspace{1em}$5 \times 8 = 40$ octet shell vacancies to conclude that this structure must violate the octet model and is thereby forced to place at least two electrons in a high-energy shell.

\subsection{Performance / duplicates}
\label{performance}
\begin{figure}[!h]

\center
\includegraphics[width=0.6\textwidth]{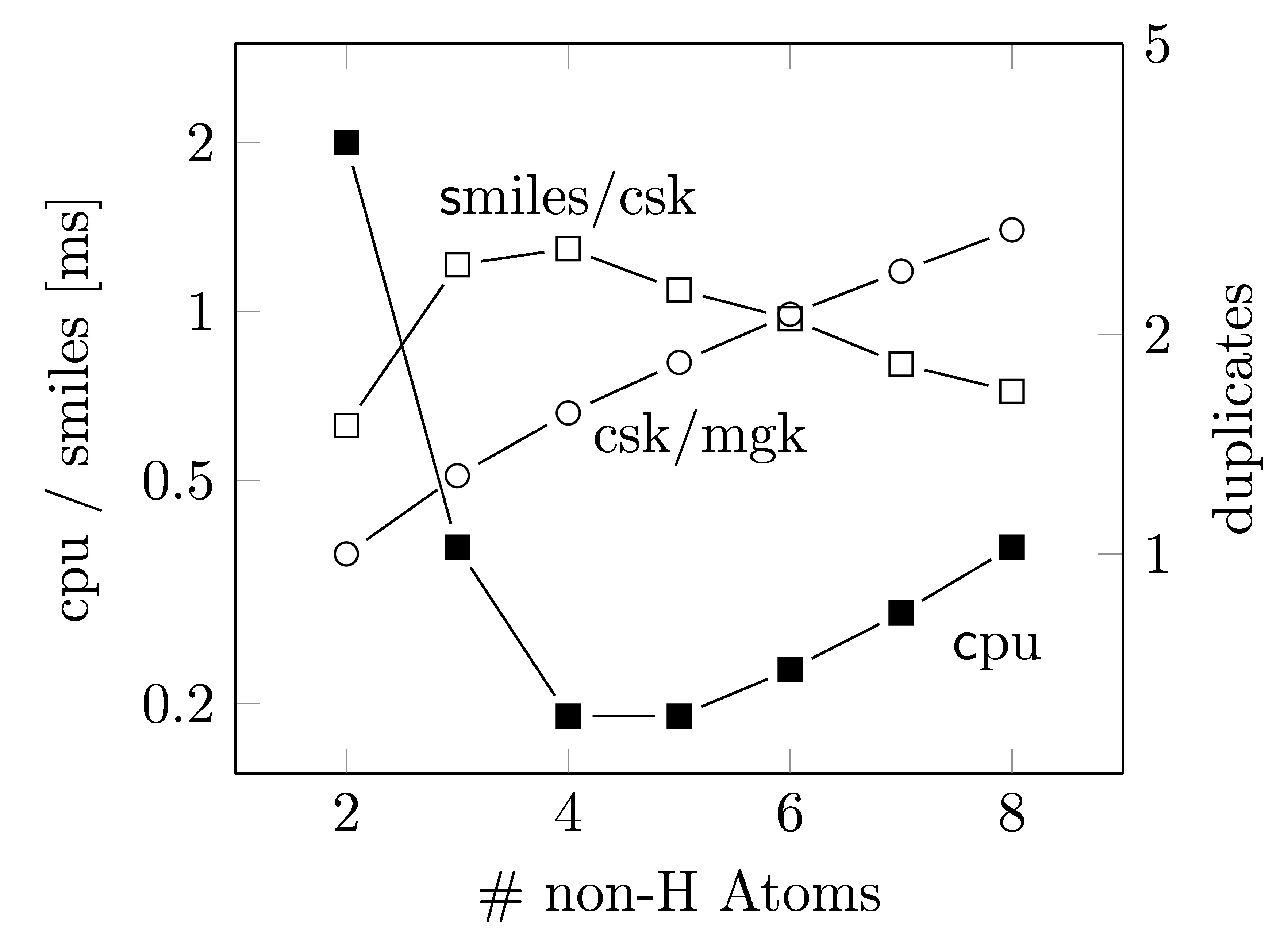}
\caption{Left-hand scale: the average CPU-time required to generate one smiles string ($\blacksquare$). Right-hand scale: two duplication factors: automorphic duplicates correspond to the number of generated smiles per unique canonical smiles ($\square$), mesomeric duplicates to the average number of unique canonical smiles per mesomeric group key ($\circ$).}
\label{FigureAnalysis}
\end{figure}

As illustrated in Figure \ref{FigureSchema}, our algorithm generates a surplus of molecules in the first three steps, from which duplicates are discarded in the last step. Such a design is efficient only if we are confident that the number of discarded duplicates is not too high. In order to check this issue we analyze the generation process in more detail in Figure \ref{FigureAnalysis}.

Figure \ref{FigureAnalysis} shows as solid squares the average CPU time\footnote{milliseconds on a 3.4 GHz Intel Core i7-3770 CPU} that is required per generated smiles string. The high value of two milliseconds for the smallest molecules is an artifact caused by the fact that it is based on generating 16 molecules only. The more reliable points start with gdb-5 and show an exponential increase. This time is dominated by the canonicalization algorithm and an exponential growth in-line with our expectations.

The relevant performance criterion is not the CPU time per smiles strings but rather the CPU time per molecule. However, this value depends strongly both on the size and structure of the generated molecules and on the exact question asked. Our generator seems to be significantly faster than OMG, as illustrated by the following example: in 6500 milliseconds CPU time we generate with our software more than 30,000 molecules matching one of the formulas \ce{C3N2OH_?}, i.e., consisting of three carbon, two nitrogen, one oxygen and any number of hydrogen atoms. For this same task OMG requires four times longer to generate
four times fewer structures. This comparison is no more than a first indication though: a fair and exact comparison would require an analysis of implementation versus algorithmic design, which is beyond the scope of this work.

The above comparison already indicates the efficiency of our algorithm. A relatively low degree of duplicate calculations is confirmed by the automorphic and mesomeric duplication
factors, included as open squares and circles in Figure
\ref{FigureAnalysis}. As an example we consider
gdb-6, where each of these factors is roughly equal to two. This means
that, in order to produce one distinct molecule (i.e., one distinct mg-key),
 we had to generate and canonicalize on average a total of four different smiles strings
that represent two distinct mesomers. Accordingly, the CPU
time per generated molecule is four times higher than the CPU time per smiles string. We find this result to be typical over the complete range of generated molecules, which confirms our design: discarding on average three out of four generated molecules seems an acceptable price to pay for a simple algorithm. This conclusion may not be valid for larger molecules, though.

The relatively constant ratio of discarded duplicates is a consequence of two canceling trends: a growing mesomeric duplication factor and a falling automorphic duplication factor. The latter trend seems surprising. It is tentatively explained by a lower average degree of symmetry among larger graphs. The idea is illustrated by the graph representing 3-methyl-hexane in Figure \ref{FigureNoOctet}(c), the smallest non-cyclic graph without symmetries. For this graph alone we generate 120,738 smiles strings in gdb-7 and among these strings not a single duplicate. Apparently such asymmetric graphs prevail among larger molecules.

\subsection{DFT calculations}
\label{dftcalculations}
In order to validate the generated molecules we performed DFT calculations for each of the 96 gdb-3 molecules. We used the program NWCHEM \cite{valiev2010nwchem} to perform these calculations, choosing the m06-2x density functional \cite{zhao2006new} and def2-TZVP basis sets \cite{weigend2005balanced,schafer1994fully} and calculating the initial three dimensional conformation that NWCHEM requires as input with Open Babel \cite{obabel}. The calculations require several days CPU time on a typical computer and this time grows strongly with size: it may seem possible to extend such calculations to the gdb-4 sets, but it is clearly out of reach to perform them on all sets.

For two cyclic molecules in our set the structure optimization did not converge, in 16 further cyclic molecules a bond was cleaved to form a linear structure. We interpret these results to mean that these structures were invalid. Stability seems correlated with electron count: with one exception the unstable rings all have a total of sixteen or less valence electrons, the stable molecules have sixteen or more valence electrons. Overall this leaves us with 78 distinct structures that are stable in a DFT calculation, more than 80\% of the generated structures.

Making the worst-case assumption that all 18 unstable structures should be considered invalid substructures in larger molecules, the sets of larger molecules should all be pruned, discarding molecules that contain one of the invalid substructures. We would expect the effect of such a pruning to be large and to grow with molecular size, but preliminary results up to gdb-9 do not show this: with $\sim\!17\%$ of the generated molecules being invalid in gdb-3, we find that the prevalence of these structures remains constant at $\sim\!17\%$ also among larger molecules.\hspace{1em}

The verification by DFT-calculations suffers from the fact that DFT itself is a model that may be incorrect. The above molecules that are confirmed by DFT do include two very unlikely ring structures. Preliminary CCSD(T) calculations that one of our reviewers recommended seem to indicate that these two structures indeed are unstable and their confirmation by DFT an artifact.

\subsection{Other chemistry and limits}
\label{otherchemistry}
\begin{figure}[!h]

\includegraphics[width=0.8\textwidth]{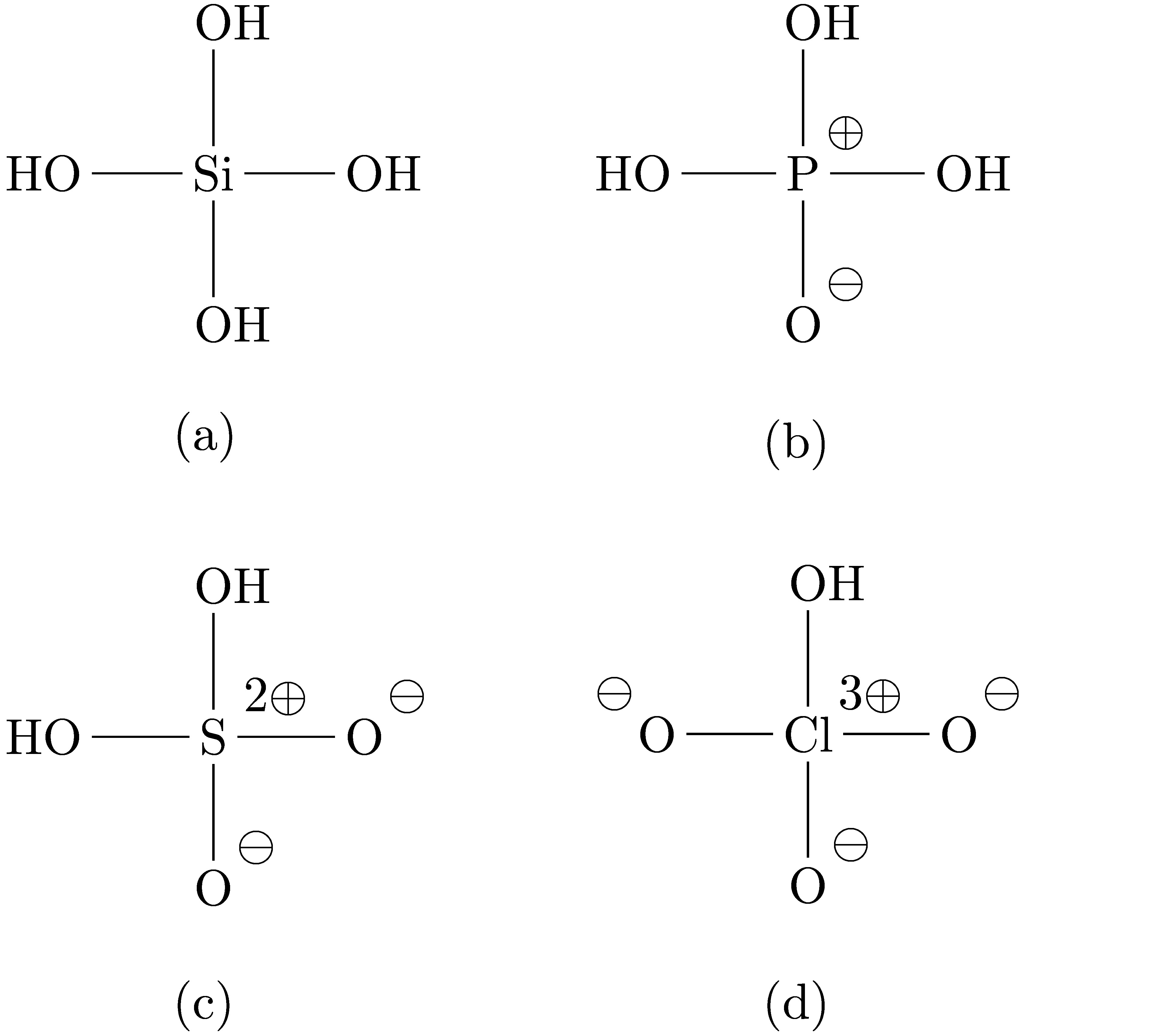}
\caption{Oxygen containing acids of third row elements as generated by our model. }

\label{FigureAcids}

\end{figure}

Our approach is not limited to the second row elements carbon, nitrogen and oxygen. To illustrate its applicability also to third-row elements we show in Figure \ref{FigureAcids} four well-known acids as they are generated by our model. For example, we generate nine isomers of \ce{H3PO4} and among those isomers also the structure shown in Figure \ref{FigureAcids}(b). Similar results are obtained for \ce{H4SiO4}, \ce{H2SO4} and \ce{HClO4}: for each of them we generate between 10 and 20 isomers that are mainly unstable peroxides and ozonides but also include the respective expected molecule.

For three of these four acids, the performance of OMG is very similar: it generates the same three correct molecules and along with each of them a number of less relevant ozonides and peroxides. The exception is perchloric acid, which is not generated by OMG, neither are chlorous and chloric acid. It seems that untypical valencies for the chlorine atom have not yet been included in the OMG-version that we use.

\begin{figure}[!h]
\includegraphics[width=0.8\textwidth]{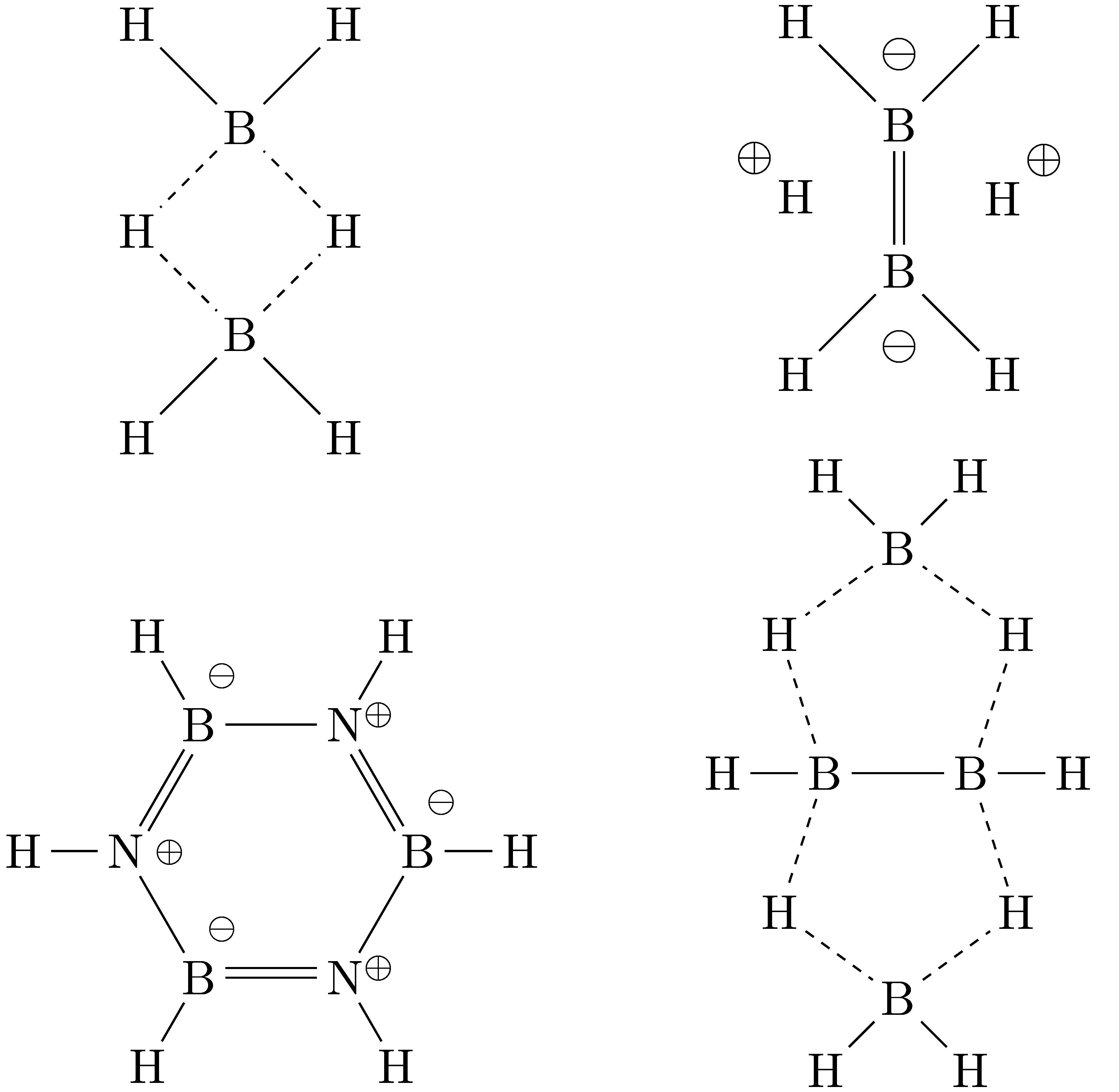}

\caption{The molecules diborane (top), borazine (bottom left) and tetraborane (right) as generated by our model. For diborane we show both the conventional depiction (left) and the generated structure that is compatible with the octet rule (right).}

\label{FigureBorane}

\end{figure}

As a further illustration we generate three examples of boron chemistry as shown in Figure \ref{FigureBorane}: diborane, borazine and tetraborane. Two of these molecules contain the three center bonds that are typical for boron chemistry. We generate these structures as shown for diborane in accordance with our octet rule model as protons that are associated with a bond between two negatively charged centers. The depicted structure is not meant to base any claims on the acidity of the proton -- it is merely a mesomeric representation that obeys the octet rule and as such provides a criterion to accept it as valid chemistry. It is straightforward to calculate the conventional depiction from our model and it is this conventional depiction that we use to display tetraborane.

It is notoriously hard to calculate the correct number of hydrogen atoms in molecules like diborane and borazine. Neither CDK, CACTUS nor Babel seem to produce correct structures here. Tetraborane is even more complicated and we don't see how we could test this structure in either one of these frameworks. We were not able to handle boron containing molecules at all with our version of OMG.

With our model we generate only the one correct structure shown in Figure \ref{FigureBorane} for the molecular formula \ce{B2H6}. For \ce{B4H10} we generate three isomers that all seem reasonable and include the correct structure as shown. For the formula \ce{B3N3H6} we generate more than 12\;000 isomers, including the expected structure as shown.

These excursions into third-row and boron chemistry illustrate the strength of our model but also show its limits.
Our model is currently limited to a coordination number of four and we can neither generate the octahedral molecule sulfur hexafluorid, nor the molecule pentaborane, in which one of the boron atoms has five neighbors.

\section{Conclusion}
\label{conclusion}
In this work we consider the graph-based generation of virtual molecules and present two improvements: We base the selection of valid chemistry on the octet rule and we exclude duplicates with the help of a newly defined mesomeric group key.

Compared to the Open Molecule Generator (OMG), which is otherwise very similar to our approach, the octet rule based selection of valid chemistry allows us to avoid a large class of invalid structures, to include many valid structures that OMG doesn't generate and to increase the performance by an order of magnitude. The versatility of the octet rule is illustrated by the correct generation of molecules containing third row elements and complex boron chemistry.

Our second improvement is the definition of a mesomeric group key (mg-key) that helps to exclude duplicates. For each generated structure we calculate the corresponding mg-key and whenever we find two or more structures that share the same mg-key, we count only one of these structures and discard the others. The need for such a filter is illustrated by the large amount of
 duplicates we find among OMG-generated structures.

The differences of our octet rule model as compared to OMG seem restricted to one specific class of molecules. We have not found a generally accepted name for this class and denote it informally as molecules having atoms with atypical valences. The formal but less-common IUPAC definition for the corresponding structures seems to be ``dipolar bonds''. A striking feature of our result is the growing prevalence of this class among larger molecules.
 Up to and including gdb-4 the difference between our generator and OMG is
almost negligible. In contrast: in gdb-9 our octet rule classifies 50\% of the OMG-generated
molecules to be incorrect and 80\% of valid molecules to be missing.

The above text and some presentations in section \ref{results} may leave the impression that we think our model is better because it generates more molecules. This is not the case. Rather, we view both generators as tools to classify smiles strings into valid and invalid molecules. Their performance should be measured in terms of four numbers: the two correct classifications valid-molecules-recognized and invalid-molecules-excluded in relation to the two incorrect classifications that are often referred to as false-negatives and false-positives. We have presented examples showing that the octet rule may be better with respect to the first three numbers, but the big unknown remains the number of false positives, i.e., the number of generated invalid molecules. Our attempt to estimate this number showed only 17\% false positives among 96 molecules generated in gdb-3, which seems good. However, we see no basis to extrapolate this result to the 500 million structures generated in gdb-9 and the high number by itself is not an advantage.

We do believe a prime strength of our model to be that it is based on one single rule: among two models of equal descriptive power, the simpler model can be expected to have more predictive power \cite{russell2013}. Based on this principle we expect a superior predictive power for our single-rule model as compared to OMG, which requires at least one valence per atom type and in addition a list of exceptions.

One prediction of our model is the high prevalence of molecules with untypical valencies. Well-known representatives of this class include, e.g., carbon monoxide, ozone, isocyanides, amine oxides and nitro compounds. The relevance of these molecules, e.g., as pharmacological compounds, is not so easily judged. The current view seems to be that these molecules are too unstable to be relevant. For example, nitro compounds are the only molecules with dipolar bonds included in GDB-11. A preliminary search seems to confirm this view, with prime applications of nitro compounds as explosives and of iso cyanides as highly reactive building blocks in chemical synthesis. On the other hand, amine oxides can be very stable, as evidenced by their widespread use as surfactants. Notably, a recent review presents heterocyclic N-oxides as an emerging class of therapeutic agents that have been successfully employed in a number of recent drug development projects \cite{mfuh2015}. A hypothesis that nitro compounds and N-oxides are the two exceptions and all other representatives are unstable is not confirmed by a search in PUBCHEM, where we found 13,400 unique small molecules that contain a positively charged nitrogen atom, of which only a very small fraction (less than 400 molecules) are N-oxides or nitro compounds. It would seem of interest to look into the use and stability of these compounds.

Whatever the outcome of a more detailed evaluation of dipolar molecules, it is very unlikely to confirm the current prediction of our generator that these molecules represent 90\% of chemical space. This leads back to the problem of false-positives and to a main direction of future research: finding additional filter criteria that help to exclude molecules that are obviously irrelevant. Examples include the ozonides we encountered when searching the oxygen containing acids, or the ring structures that were unstable in the DFT structure optimization. Such an additional filter could be based on the torsional and steric forces that arise from a three-dimensional model of a molecule.

Also an exhaustive enumeration of all tautomers seems a worthwhile future improvement. This seems straightforward to implement and would allow a clearer insight into the structure of chemical space. Furthermore, extensions to charged molecules, to coordination numbers higher than four and/or to molecules
with occupied d-orbitals seem possible. An obvious improvement of our mesomeric group keys would be the correct handling of stereo-isomerism. The recognition of asymmetric C-atoms is already available as a component of the canonicalization algorithm and could be included using the notation that is known in smiles strings.

\bibliographystyle{plain}      % basic style, author-year citations
\bibliography{IsraelsMaassHamaekers2017}

\end{document}